\documentclass[english,aps,showpacs,prb,twocolumn,floatfix,groupedaddress,a4paper]{revtex4-1}
\usepackage{amsmath}
\usepackage{graphicx}
\usepackage{amssymb}
\usepackage[usenames,dvipsnames]{color}
\usepackage{hyperref}
\usepackage{natbib}
\usepackage{dsfont}
\usepackage{units}
\usepackage{placeins}
\usepackage{bm}
\usepackage{ulem}
\usepackage{dcolumn}
\usepackage{braket}
\usepackage[mathscr]{eucal}

\makeatletter
\usepackage{babel}
\makeatother

\newcommand{\pdag}{{\phantom{\dagger}}}

\begin{document}
\title{Strongly interacting holes in Ge/Si nanowires}
\date{\today}
\author{Franziska Maier}
\affiliation{Department of Physics, University of Basel, Klingelbergstrasse 82, CH-4056 Basel, Switzerland}
\author{Tobias Meng}
\affiliation{Department of Physics, University of Basel, Klingelbergstrasse 82, CH-4056 Basel, Switzerland}
\author{Daniel Loss}
\affiliation{Department of Physics, University of Basel, Klingelbergstrasse 82, CH-4056 Basel, Switzerland}
\begin{abstract}
We consider holes confined to Ge/Si core/shell nanowires subject to strong Rashba spin-orbit interaction and screened Coulomb interaction. 
Such wires can, for instance, serve as host systems for Majorana bound states.
Starting from a microscopic model, we find that the Coulomb interaction strongly influences the properties of experimentally realistic wires. To show this, a Luttinger liquid description is derived based on a renormalization group analysis. This description in turn allows us to calculate the scaling exponents of various correlation functions as a function of the microscopic system parameters. It furthermore permits us to investigate the effect of Coulomb interaction on a small magnetic field, which opens a strongly anisotropic partial gap.
\end{abstract}
\maketitle
%
%
%
%
%
%
\section{Introduction}
In the past decades, semiconductor nanowires (NWs) have proven to be a versatile platform for the engineering of nanoscale systems, both as intrinsically one-dimensional (1D) channels, and as hosts for NW quantum dots (QDs). So far, NWs have predominantly been grown using III-V compounds, which can be operated both in the electron regime,\cite{Doh2005,Fasth2007,Nilsson2009,NadjPerge2010,Schroer2011,NadjPerge2012,Petersson2012,vandenBerg2012} and the hole regime. \cite{Pribiag2013} Recently, a new class of  NWs, made of a cylindrical Ge core and a Si shell,\cite{Lauhon2002,Lu2005,Xiang2006,Xiang2006a,Hu2007,Roddaro2007,Roddaro2008,Yan2011,Nah2012,Hu2012} and ultrathin triangular Ge NWs on a Si substrate,\cite{Zhang2012} have emerged as promising alternatives to III-V NWs. The core/shell NWs can be grown with core diameters of $5-100\mbox{ nm}$, and shell thicknesses of $1-10\mbox{ nm}$. Inside the core, a 1D hole gas accumulates,\cite{Lu2005,Park2010} and the $p$-wave symmetry of the hole Bloch states results 
in an  
unusually large and tunable Rashba-type spin-orbit interaction (SOI).\cite{Kloeffel2011} Applying a magnetic field allows one to access a helical regime \cite{Kloeffel2011} susceptible to the formation of Majorana zero-energy bound states (MBS) when the NW is proximity coupled to an $s$-wave superconductor.\cite{Alicea2012}
Finally, when grown nuclear spin free, these systems have significantly reduced hyperfine induced decoherence effects. Experimentally, high mobilities,\cite{Xiang2006,Nah2012} long mean free paths,\cite{Lu2005} proximity-induced superconductivity,\cite{Xiang2006a} and signatures of the tunable Rashba SOI \cite{Hao2010} have been identified. Longitudinal confinement has been demonstrated to create tunable single and double QDs,\cite{Hu2007} with anisotropic and confinement dependent $g$ factors,\cite{Roddaro2007,Roddaro2008} short SOI lengths,\cite{Higginbotham2014a}as well as long singlet-triplet relaxation times \cite{Hu2012} and hole spin coherence times.\cite{Higginbotham2014b} Holes confined to such QDs have 
furthermore been predicted to exhibit strongly anisotropic, tunable $g$ factors and long spin phonon relaxation times,\cite{Maier2013} and have been proposed as a platform for quantum information processing.\cite{Kloeffel2013}

In this paper, the effects of hole-hole interactions, and their Luttinger liquid description in Ge/Si core/shell NWs are, to the best of our knowledge, addressed and quantified for the first time based on a concrete microscopic model. We focus on the single subband regime most relevant for the emergence of MBS. After explicitly evaluating the interaction matrix elements for a realistic geometry,  we derive the Luttinger liquid description of the NW, and calculate the interaction dependent scaling exponents of various correlation functions for our microscopic model. The scaling exponents show a weak dependence on the magnitude of an applied electric field, which tunes the SOI strength. This is contrasted by a strong dependence on the NW parameters. The exponents differ substantially from their non-interacting value, thus revealing rather strong interaction effects. As an example for experimental implications of Luttinger liquid 
physics beyond the scaling of correlation functions, we finally analyze 
the renormalization of the partial gap around zero momentum resulting from an applied magnetic field. This partial gap precisely corresponds to the helical regime susceptible to the formation of MBS in a superconducting hybrid device.\cite{Alicea2012} We find that hole-hole interactions lead to a sizable enhancement of the gap (thus implying more stable MBS in an interacting system), which is furthermore strongly anisotropic.

The outline of this paper is as follows. In Sec.~\ref{sec:Model} we introduce the effective 1D Hamiltonians describing holes in Ge/Si NWs interacting via Coulomb repulsion and distill an effective lowest-energy Hamiltonian. We bosonize the latter in Sec.~\ref{sec:bosonization} and, in Sec.~\ref{sec:exps_corrfuncs}, analyze the exponents of the correlation functions regarding the dependence on the applied electric field and NW parameters. In Sec.~\ref{sec:renormaliz_gap}, we examine the partial gap opened by an external magnetic field and its dependence on the electric field and the direction of the magnetic field. 
For technical details we refer to the Appendixes. 

%
%
%
%
%
\section{Model \label{sec:Model}}
\subsection{1D hole Hamiltonian}
As a first step, we derive an effective theory for the single subband regime of a Ge/Si core/shell NW in the presence of Coulomb interactions. Our starting point is a more complex model \cite{Kloeffel2011} for a NW with core (shell) radius $R$ ($R_s$) aligned with the $z$ axis of the coordinate system, and exposed to an electric field perpendicular to the NW axis, $\bm{E} = E_{\perp}(\cos\varphi_E, \sin\varphi_E, 0)$. A possibly applied magnetic field will be added in a later step. The non-interacting part of this setup is well described by an effective quasi-1D Hamiltonian $H_0=\int d z \,\mathscr{H}_0$ with 
\begin{align}
\mathscr{H}_0 &= \Psi^{\dagger}(z) \left[\mathscr{H}_{\text{LK}} + \mathscr{H}_\text{strain} +\mathscr{H}_{\text{R}}+\mathscr{H}_{\text{DR}}-\mu\right] \Psi(z)~,\label{eq:Hamiltonian4x4RashbaA}
\end{align}
where $\mu$ denotes the chemical potential, and with $\mathscr{H}_0$ being written in the basis $\{\Psi_{g_+}(z), \Psi_{g_-}(z),\Psi_{e_+}(z), \Psi_{e_-}(z)\}$. The indices $g_{\pm},e_{\pm}$ comprise the band $(g,e)$ and pseudospin $(+,-)$ labels, and the annihilation operators are given by $\Psi_{i}(z) = \sum_{k_z} e^{i k_z z} c_{i,k_z}$, with $c_{i,k_z}$ being the annihilation operator of a hole state $i$ with momentum $k_z$ along the NW. 
The Luttinger-Kohn and strain Hamiltonian densities read $\mathscr{H}_{\text{LK}} + \mathscr{H}_\text{strain} = A_{+}(k_z) + A_{-}(k_z) \tau_z +  C k_z \tau_y \sigma_x$, with $\tau_i$ and $\sigma_i$ being the Pauli matrices acting in the band and pseudo-spin space, respectively. 
Here, $A_{\pm}(k_z, \eta) \equiv \hbar^2 k_z^2(m_g^{-1} \pm m_e^{-1})/4\mbox{ $\pm$ } \Delta/2$, with Planck's constant $\hbar$, and with effective masses $m_g\simeq m_0/(\gamma_1+2\gamma_s)$ and $m_e = m_0/(\gamma_1+\gamma_s)$.
The bare electron mass is denoted by $m_0$, and $\gamma_1$ and $\gamma_s$ are the Luttinger parameters in spherical approximation. 
For Ge, $\gamma_1=13.35$ and $\gamma_s=5.11$.\cite{Lawaetz1971}
The level splitting between the $g_{\pm}$ and $e_{\pm}$ states is $\Delta \equiv \Delta_{\rm LK} + \Delta_{\rm strain}(\eta)$ with relative shell thickness $\eta \equiv (R_s - R)/R$, confinement induced $\Delta_{\text{LK}}=0.73 \hbar^2/(m_0 R^2)$ and the strain dependent splitting $\Delta_{\text{strain}}(\eta)\simeq0-30\mbox{ meV}$. 
The off-diagonal coupling with coupling constant $C=7.26 \hbar^2/(m_0 R)$ is a direct consequence of the strong atomic level SOI.
The direct Rashba SOI, $\mathscr{H}_{\text{DR}} = e U E_{\perp}(\tau_x \sigma_z\cos\varphi_E  - \tau_y\sin\varphi_E  )$, where $U=0.15 R$, results from direct, dipolar coupling of $\bm{E}$ to the charge of the hole. 
The conventional Rashba SOI reads $\mathscr{H}_{\text{R}}= \alpha_R E_{\perp} [S(\tau_x \sigma_z \cos\varphi_E  - \tau_y\sin\varphi_E )+ B_{+}(k_z) + B_{-}(k_z) \tau_z ]$,
with $B_{\pm}(k_z) \equiv k_z T(\sigma_x \sin\varphi_E  +  \sigma_y\cos\varphi_E)/2\mp3 k_z ( \sigma_x \sin\varphi_E -  \sigma_y \cos\varphi_E)/8$, where $T=0.98$, $S=0.36/R$, and $\alpha_R=-0.4 \mbox{ nm}^2 e$ with elementary charge $e$.
Note that $e U/(\alpha_R S) \simeq -1.1 R^2\mathrm{nm}^{-2}$, hence $\mathscr{H}_{\text{DR}}$ dominates $\mathscr{H}_{\text{R}}$ by one to two orders of magnitude for  $R=5-10\mbox{ nm}$. 
Diagonalizing the full $(4 \times 4)$ matrix Hamiltonian $H_0$ yields the eigenenergies $E_{g_{+}'}$, $E_{g_{-}'}$, $E_{e_{+}'}$ and $E_{e_{-}'}$. 
The  associated annihilation operators are $\Psi_{g_+'}(z)$, $\Psi_{g_-'}(z)$, $\Psi_{e_+'}(z)$, and  $\Psi_{e_-'}(z)$,  which are linear combinations of the original annihilation operators introduced below Eq.~(\ref{eq:Hamiltonian4x4RashbaA}). 
The coefficients of the linear combinations depend strongly on the NW parameters $R$ and $\Delta$, and both magnitude and direction of $\bm{E}$. In the following, we assume the chemical potential to be placed below the bottom of the upper bands $e_{\pm}'$, and therefore focus on the low-energy Hamiltonian in the subspace spanned by $\{\Psi_{g_+'}(z), \Psi_{g_-'}(z)\}$. 

\subsection{Coulomb Interaction}
Next, we generalize this Hamiltonian to the interacting case. For our concrete microscopic model, we assume the holes to interact via Coulomb repulsion, and take the latter to be screened by mirror charges in the nearby gates. The associated potential for a hole located at $\bm{r}$ interacting with a hole located at $\bm{r}'$ in the presence of a mirror charge at $\bm{r}_{\text{mc}}$ is given by 
\begin{equation}
V(\bm{r},\bm{r}',\bm{r}_{\text{mc}}) = \frac{e^2}{4 \pi \varepsilon_0 \varepsilon_r}\left[\frac{1}{|\bm{r}-\bm{r}'|}-\frac{1}{|\bm{r}-\bm{r}_{\text{mc}}|}\right],
\end{equation}
with vacuum permittivity $\varepsilon_0$, and relative permittivity $\varepsilon_r$.
For Ge, $\varepsilon_{r} \approx 16$.\cite{[{See e.g. }]DAltroy1956} In the initial $(4\times4)$ basis, the interaction Hamiltonian thus reads
\begin{align}
H_{c} =&\frac{1}{2}\sum_{ijkl}\!\int\!\!\!\!\int \mathrm{d}z\mathrm{d}z' \Psi_{i}^{\dagger}(z)\Psi_{j}^{\dagger}(z') \nonumber\\
&\times \left[\int\mathrm{d}q V_{1D}^{ijkl}(q) e^{i q [z-z']}\right]\Psi_{k}(z')\Psi_{l}(z), \label{eq:Hinteract}
\end{align}
with $i,j,k,l = g_{\pm}, e_{\pm}$, and $q$ being the wavevector along the NW. The interaction matrix elements $V_{1D}^{ijkl}(q) $ of $H_c$ are obtained by integrating out the transverse part of $V(\bm{r},\bm{r}',\bm{r}_{\text{mc}})$ using the three-dimensional wavefunctions of holes in Ge/Si NWs derived in Ref.~[\onlinecite{Kloeffel2011}]. 
A more detailed sketch of this calculation is given in Appendix \ref{app:calc_CoulMatel}.
Finally, we project the full $(4\times4)$ interaction Hamiltonian $H_c$ onto the diagonalized low energy subspace $E_{g_\pm}'$, thus arriving at the interacting effective model for the single subband regime of a Ge/Si core/shell NW.

\begin{figure}[t]
 \includegraphics[width = 0.8\columnwidth]{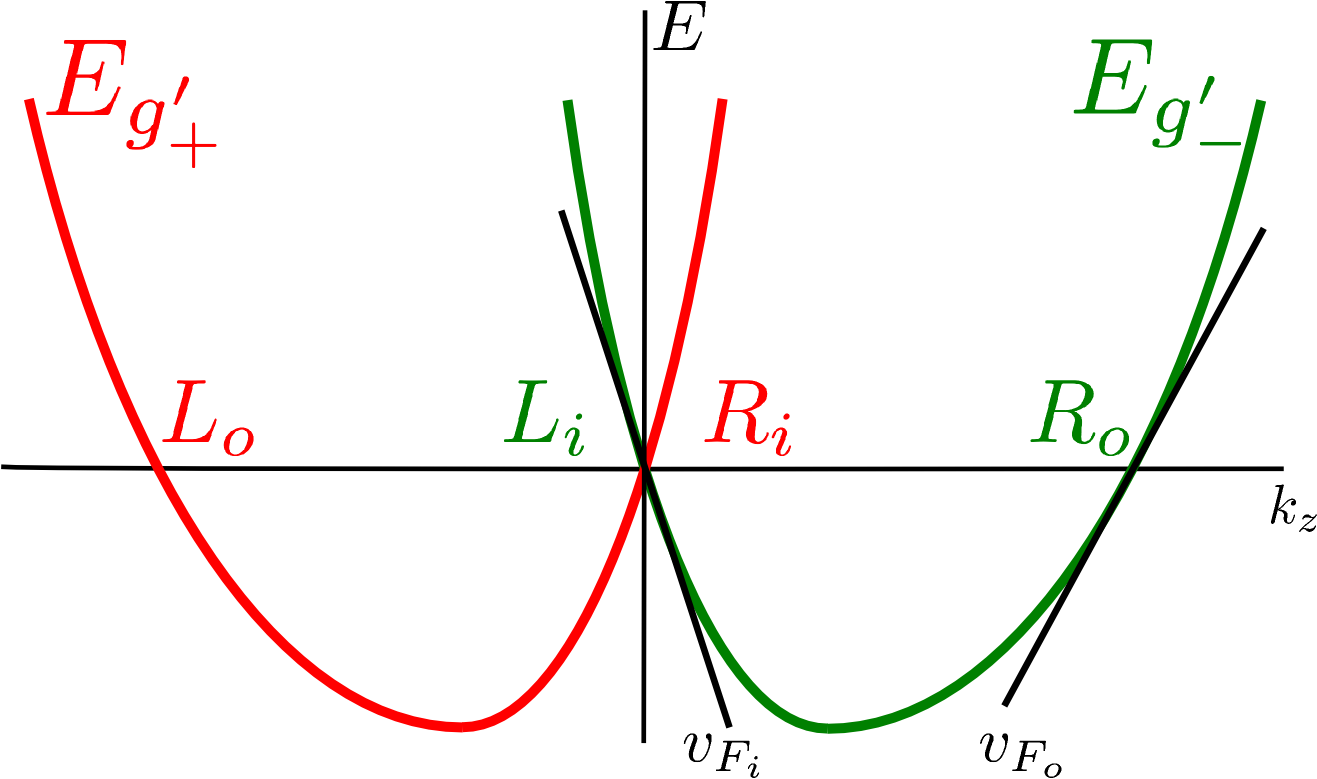}
 \caption{Sketch of the low-energy dispersions $E_{g_{+}'}$ and $E_{g_{-}'}$ as functions of the momentum $k_z$ along the NW for a finite field $E_{\perp}$.
The Fermi velocities $v_{F_i}$ and $v_{F_o}$ for the inner and outer modes differ in the general case, \textit{i.e.}~$v_{F_i}\neq v_{F_o}$.
 \label{fig:Rashbasystem}}
\end{figure}
%
%
%
%
%
\section{Bosonization \label{sec:bosonization}}
The low energy excitations of this interacting 1D system are given by collective bosonic density waves rather than individual fermionic quasiparticles.\cite{Giamarchi2003} To distill the related Luttinger liquid Hamiltonian, we linearize the non-interacting part of the spectrum, depicted in Fig.~\ref{fig:Rashbasystem}, around the Fermi points. In this process, we only retain low energy excitations by introducing a momentum cutoff $\hbar/\alpha$ relative to the Fermi points, where $\alpha$ denotes the short distance cutoff length. Because of the SOI, the pseudospin bands $g_+'$ and $g_-'$ are split in momentum space. We decompose the operators $\Psi_{g_\pm'}(z)$ into right ($R_j$) and left ($L_j$) moving modes associated with the low energy excitations close to the inner ($j=i$) and outer ($j=o$) Fermi points, $\Psi_{g_+'}(z) \simeq R_i(z) e^{i k_{F_i}z} + L_o(z) e^{-i k_{F_o}z}$ and $\Psi_{g_-'}(z) 
\simeq R_o(z) e^{i k_{F_o}z}+L_i(z) e^{-i k_{F_i}z}$, where the inner and outer Fermi wavenumbers are $k_{F_{i,o}}$. The slopes of the spectrum at these points define the Fermi velocities $v_{F_{i}}$ and $v_{F_{o}}$. These differ because the admixing of the higher energy bands $e_{\pm}$ renders the bands $g_{\pm}'$ non-parabolic. Since we are eventually interested in the renormalization of the partial gap opened by a small magnetic field, we furthermore choose the chemical potential to be pinned to the crossing point of $E_{g_{+}'}$ and $E_{g_{-}'}$. We emphasize, however, that our model is valid for arbitrary values of $\mu$, with the exception of $\mu$ being close to the bottom of the band, where the non-linearity of the spectrum becomes important for the low-energy excitations.

While the projection of the non-interacting Hamiltonian $H_0$ on the low-energy modes $R_{i,o}$ and $L_{i,o}$ simply reads $H_0 \approx \int\mathrm{d}z \sum_{s=i,o} v_{F_s}(R_s^{\dagger} \partial_z R_s-L_s^{\dagger} \partial_z L_s)$, the interaction Hamiltonian $H_c$ demands a more careful treatment. We project $H_c$ on the low-energy modes $R_{i,o}$ and $L_{i,o}$, thereby dropping rapidly oscillating terms, and classify the remaining $V_{1D}^{ijkl}(q)$ according to the standard $g$-ology.\cite{Giamarchi2003} This translates $V_{1D}^{ijkl}(q)$ to the interaction matrix elements $g_{n_f}$ with indices $n=1,2,4$, and $f=i,o,io$, which couple only inner ($i$), only outer ($o$), or inner and outer ($io$) modes. {
Note that we observe several matrix elements corresponding to $g_1$ processes coupling the inner and outer modes, we label them $g_{1_{ioj}}$, $j=1,2,3,4$, in order of appearance. 
With these definitions, the projection of $H_c$ reads $H_c=\int \mathrm{d}z\,(\mathscr{H}_1 + \mathscr{H}_2 + \mathscr{H}_4)$, with
\begin{align}
\mathscr{H}_2 =& g_{2_i}\,\rho_{R_i}  \rho_{L_i} + g_{2_o}\, \rho_{L_o}  \rho_{R_o}+ g_{2_{io}}  \left(\rho_{L_i} \rho_{R_o}  + \rho_{L_o} \rho_{R_i}\right),\label{eq:H2}\\ 
\mathscr{H}_4=&\frac{g_{4_i}}{2} \left(\rho_{R_i}^2 + \rho_{L_i}^2 \right)+ \frac{g_{4_o}}{2} \left(\rho_{R_o}^2 + \rho_{L_o}^2 \right)\label{eq:H4}\\
&+  g_{4_{io}}\left( \rho_{R_i} \rho_{R_o} + \rho_{L_i} \rho_{L_o} \right)\nonumber,\\
\mathscr{H}_1 =&  2 g_I( R_i^{\dagger} L_i^{\dagger} L_o R_o +  
\mathrm{h.c.})
\label{eq:H1}\\
&- g_{1_{io1}}\left( \rho_{R_i} \rho_{L_o}+  \rho_{R_o} \rho_{L_i} \right)- g_{1_{io2}}\left( \rho_{L_o} \rho_{L_i} +  \rho_{R_o} \rho_{R_i}\right).\nonumber
\end{align}
where $g_I=(g_{1_{io3}}-g_{1_{io4}})/2$, and with $\rho_{r_j} = r_{j}^\dagger r_{j}^\pdag$ ($r=R,L$). 
Note that we have dropped the terms proportional to $g_{1_i}$ and $g_{1_o}$ because their matrix elements vanish, while we obtain $g_{2_f} = g_{4_f}$, $g_{1_{io1}}=g_{1_{io4}}$, and $g_{1_{io2}} = -g_{1_{io3}}$. 

We thus find that as usual,\cite{Giamarchi2003} the Coulomb repulsion gives rise to several terms proportional to squares of the fermionic densities $\rho_{rj}$, plus the term proportional to $g_I$. 
We bosonize these interaction terms by expressing the fermionic single-particle operators as $r_s = U_{r,s}  e^{-i/2[(1+s)(r \phi_i-\theta_i)+(1-s)(r\phi_o-\theta_o)]}/\sqrt{2 \pi \alpha}$, where $r=R,L\equiv+1,-1$ 
labels the chirality, and $s=i,o\equiv+1,-1$ denotes the inner/outer-pseudospin, while $U_{r,s}$ are Klein factors (unessential for our discussion). The bosonic fields $\phi_{s}$ relate to the integrated density of $s=i,o$ particles, while the canonically conjugate fields $\theta_s$ are 
proportional to their current. In terms of the bosonic fields, the Hamiltonian takes the form $H= \int \mathrm{d}z \,(\Psi_\phi^\dagger \mathscr{H}_\phi\Psi_\phi +  \Psi_\theta^\dagger \mathscr{H}_\theta\Psi_\theta)/2\pi +  \int \mathrm{d}z \, g_I \cos[2(\theta_o-\theta_i)]/(\pi \alpha)^2$, where $\Psi_\phi = (\partial_z\phi_o, \partial_z\phi_i)^T$, $\Psi_\theta = (\partial_z\theta_o, \partial_z\theta_i)^T$, and
\begin{subequations}
\begin{align}
\mathscr{H}_{\phi}&=\begin{pmatrix}
v_{F_o}+\frac{g_{4_o}+g_{2_o}}{2\pi}&\frac{g_{4_{io}}+g_{2_{io}}-g_{1_{io1}}-g_{1_{io2}}}{2\pi}\\
\frac{g_{4_{io}}+g_{2_{io}}-g_{1_{io1}}-g_{1_{io2}}}{2\pi}& v_{F_i}+\frac{g_{4_i}+g_{2_i}}{2\pi}\end{pmatrix}~,\label{eq:mat1}\\
\mathscr{H}_{\theta}&=\begin{pmatrix}
v_{F_o}+\frac{g_{4_o}-g_{2_o}}{2\pi}&\frac{g_{4_{io}}-g_{2_{io}}+g_{1_{io1}}-g_{1_{io2}}}{2\pi}\\
\frac{g_{4_{io}}-g_{2_{io}}+g_{1_{io1}}-g_{1_{io2}}}{2\pi}& v_{F_i}+\frac{g_{4_i}-g_{2_i}}{2\pi}\end{pmatrix}~. \label{eq:mat2}
\end{align}\label{eq:bos_mat}
\end{subequations}
The quadratic sector of the Hamiltonian can be diagonalized by a canonical transformation, resulting in effective low-energy degrees of freedom with velocities $u_p$ and $u_m$ (see Appendix \ref{sec:app_op}), while the sine-Gordon term $\sim g_I$ is analyzed using a standard perturbative renormalization group (RG) approach.\cite{Giamarchi2003} Because we choose to fix the chemical potential at the crossing point of $E_{g_{+}'}$ and $E_{g_{-}'}$, our calculation is restricted to sufficiently large electric fields $E_\perp$ such that  $g_I$ can be treated as a perturbation ($2 g_I/(u_p+u_m)\ll 1$).  For smaller $E_\perp$ one of the velocities, $u_m$, vanishes, and the dimensionless $g_{I}$ becomes non-perturbatively large. With this restriction in mind, we find that $g_I$ is an RG irrelevant perturbation in the regime described by our calculation.
\section{Exponents of the correlation functions\label{sec:exps_corrfuncs}}
After integrating the RG flow of $g_I$ to weak coupling, we evaluate various correlation functions $\langle O_{j}^{\dagger}(r)O_{j}(0)\rangle$ for the charge and spin density waves ($j=\text{CDW}, \text{SDW}$), and the singlet and triplet superconducting fluctuations ($j=\text{SS}, \text{TS}$)\cite{Giamarchi2003}  (see Appendix \ref{sec:app_op}),  where the spin is the pseudospin distinguishing the bands ${g_\pm}'$ (as detailed in Appendix \ref{sec:app_op}, the correlation functions for SDW$_{x,y}$, SS, and TS$_z$ comprise two terms with slightly different exponents).
The scaling exponents of these correlation functions are depicted in Fig.~\ref{fig:exponents_Edependence} as functions of the applied field $E_{\perp}$ for one concrete set of NW parameters, and exhibit only a weak dependence on $E_\perp$ (the same is found for other NW parameters). In Fig.~\ref{fig:exponents_SG}, we furthermore plot the scaling exponents for 12 concrete sets of system parameters at a fixed field $E_{\perp}$. In general, our microscopic model predicts that the exponents of the correlation functions show a strong dependence on the microscopic NW parameters $R$ and $\Delta$, determined by the core and shell radii. The scaling exponents differ substantially from $2$, their non-interacting value, thus indicating strong interaction effects in Ge/Si core/shell NWs. Exponents differing the most from $2$ are found for the NW parameter set with the smallest $R$, indicating that thin NWs show the strongest interactions. 
We note that when the field $E_\perp$ is tuned to sufficiently small values such that the system is pushed outside the perturbative regime, the bosonic RG calculation exhibits a Wentzel-Bardeen singularity.\cite{Wentzel1951, Bardeen1951, Loss1994,Martin1995} As a crosscheck for the absence of singularities in the regime well-described by our calculation, we have performed a fermionic one-loop RG analysis,\cite{Muttalib1986, Schulz2009} which reproduces the non-singular behavior of the scaling exponents in the perturbative regime. In the $E_\perp$-range where also the fermionic calculation is not valid, the presence or absence of a singularity in the one-loop calculation depends strongly on the chosen NW parameters. For more details, we refer the reader to Appendix \ref{app:divergence_exps}.

\begin{figure}[t]
\includegraphics[width = \columnwidth]{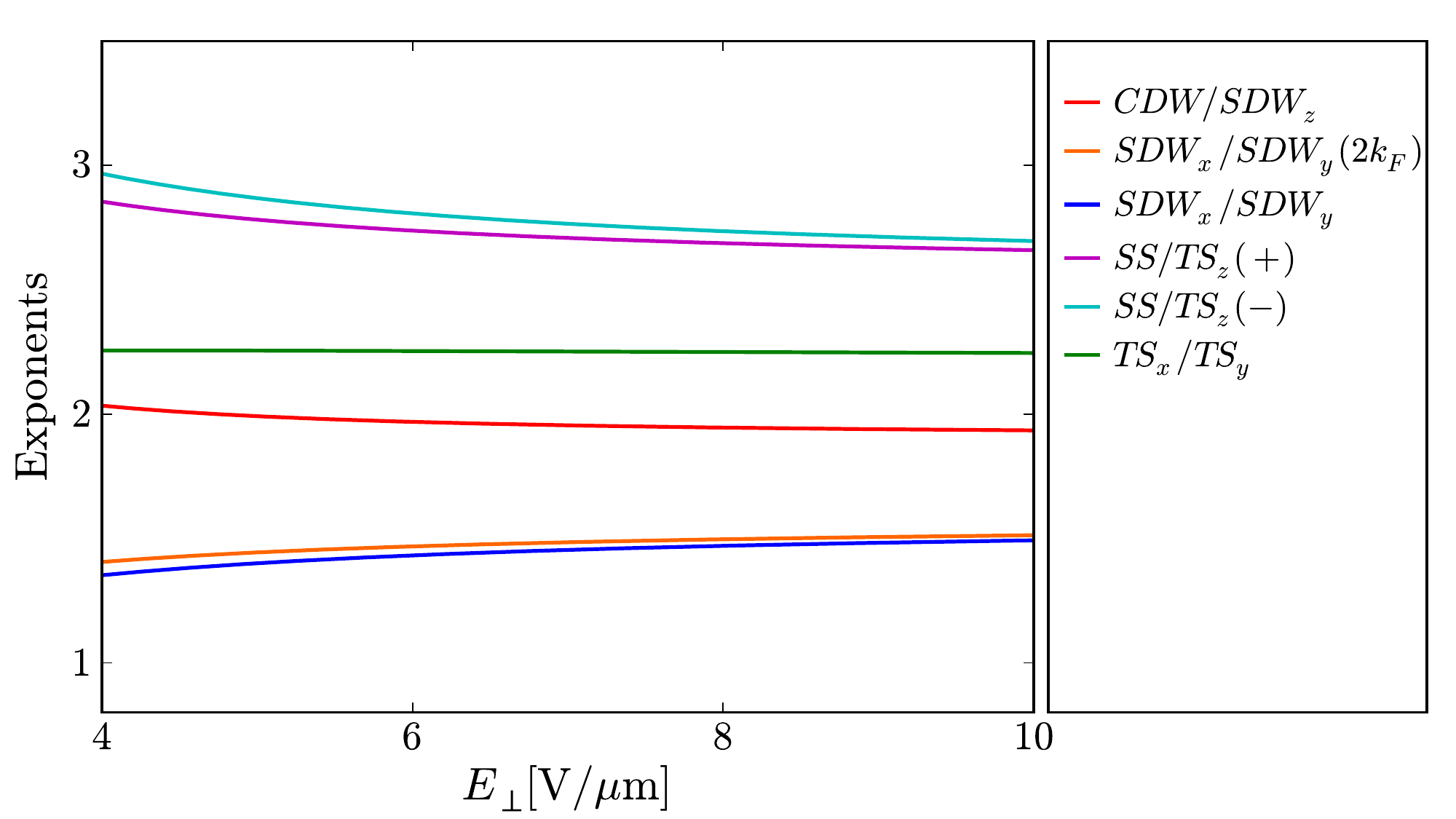}
\caption{The exponents of the correlation functions as functions of $E_{\perp}$ for $R = 10\mbox{ nm}$ and $\Delta = 8\mbox{ meV}$  in the regime where $g_I$ can be treated as perturbation. We fix $\varphi_E = 3\pi/2$ and $\bm{r}_{\text{mc}} = (0,|\bm{r}_{\text{mc}}|,0)$ with $|\bm{r}_{\text{mc}}|=100\mbox{ nm}$.
\label{fig:exponents_Edependence}}
\end{figure}

\begin{figure}[t]
\includegraphics[width = \columnwidth]{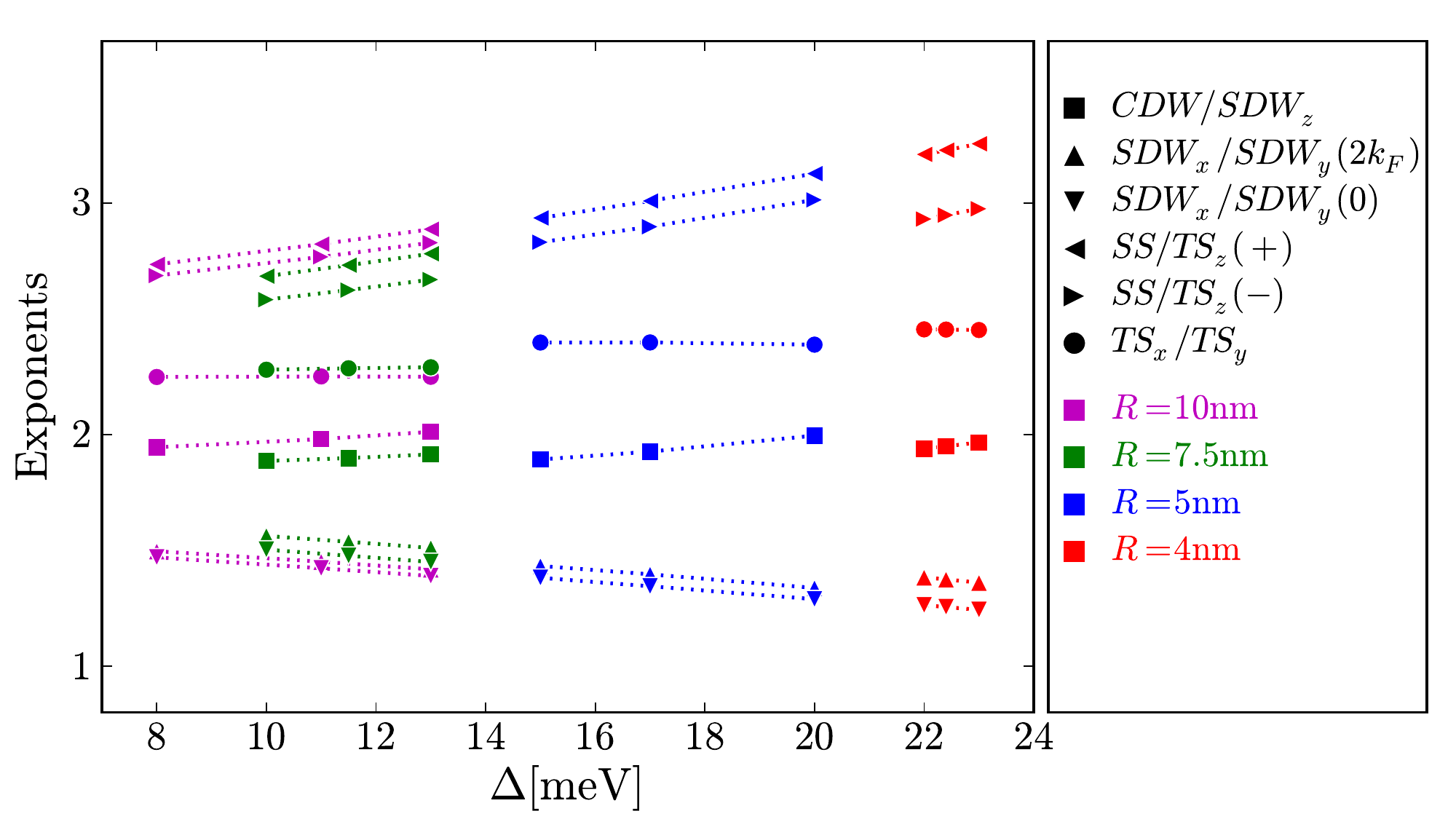}
\caption{
The exponents of the correlation functions as functions of $\Delta$ for four different core radii, $R=4\mbox{ nm}$ (red), $R=5\mbox{ nm}$ (blue), $R=7.5\mbox{ nm}$ (green), and $R=10\mbox{ nm}$ (magenta), where $E_{\perp} = 8 \mbox{ V/$\mu$m}$ at fixed $\varphi_E = 3\pi/2$ and $\bm{r}_{\text{mc}} = (0,|\bm{r}_{\text{mc}}|,0)$ with $|\bm{r}_{\text{mc}}|=100\mbox{ nm}$. For a fixed $R$ and increasing $\Delta$, all exponents besides the ones for the CDW/SDW$_z$ and the TS$_x$/TS$_y$ become increasingly different from their non-interacting value, $2$. 
The dotted lines are guide to the eyes. 
\label{fig:exponents_SG}}
\end{figure}
%
%
%
%
%
%
\section{Renormalization of the partial gap\label{sec:renormaliz_gap}}
As a final example for interaction effects in Ge/Si core/shell NWs, we now turn to the rescaling of the gap opened by a small magnetic field $\bm{B} = (B_{\perp}\cos\varphi_E, B_{\perp}\sin\varphi_E,B_z)$. This gap, giving rise to the helical regime susceptible to the formation of MBS, is known to be enlarged by Coulomb interaction in an electronic Rashba NW.\cite{Braunecker2009b,Braunecker2010} 
To analyze this effect in our concrete microscopic model with hole-hole interactions, we first introduce the magnetic field Hamiltonian density $\mathscr{H}_B = \mathscr{H}_{B,Z} + \mathscr{H}_{B,\text{orb}}$ in the original fermionic $(4\times4)$ basis of $\mathscr{H}_0$ with $\mathscr{H}_{B,Z} = [C_{+}+C_{-}\tau_z]\sigma_z +[D_{+}+D_{-}\tau_z] \sigma_x \cos\varphi_E-[D_{-}+D_{+}\tau_z]\sigma_y \sin\varphi_E$ and $\mathscr{H}_{B,\text{orb}} = F_z \tau_x \sigma_y+ F_{\perp}[\tau_y\cos\varphi_E  + \tau_x\sigma_z \sin\varphi_E ]$. Here, $C_{\pm} = \mu_B B_z (F\pm G)/2$, $D_{\pm} = \mu_B B_{\perp}(K\pm M)/2$, $F_z = \mu_B B_z D k_z$ and $F_{\perp} = \mu_
B 
B_{\perp} L k_z$ with $F = 1.56$, $G = -0.06$, $K = 2.89$, $M = 2.56$, $D = 2.38 R$ and $L = 8.04 R$.\cite{Kloeffel2011} We focus on a magnetic field $\bm{B}$ in the plane defined by $\bm{E}$ and the NW axis since a field perpendicular to this plane does not give rise to the helical regime relevant for MBS, but rather to a spin-polarized state. To bring this field to its bosonized form, we first transform $\mathscr{H}_B$ according to the unitary transformation that diagonalizes the fermionic $(4\times4)$ Hamiltonian, and then project it to the lower bands. This yields a Hamiltonian density of the form $\mathscr{H}_{B}' = \mu_B \left[g_z B_z\sigma_x + g_{\perp}B_{\perp}\sigma_y\right]$, with effective $g$ factors $g_z$ and $g_{\perp}$, and where $\sigma_{x,y,z}$ acts on the pseudospin distinguishing $g_\pm'$.
We finally bosonize $\mathscr{H}_{B}'$, and obtain
\begin{equation}
\mathscr{H}_B' =  \frac{1}{\pi \alpha}\Delta_Z(\vartheta_B) \cos[2 \phi_i-\varphi_B],
\end{equation}
with $\Delta_Z(\vartheta_B) = \mu_B (g_z^2 B^2\cos^2\vartheta_B + g_{\perp}^2 B^2\sin^2\vartheta_B)^{1/2}$, $B_{\perp} = B\sin\vartheta_B$, $B_z=B\cos\vartheta_B$, and $\tan\varphi_B = g_z B_z/(g_{\perp}B_{\perp})$. This sine-Gordon term obeys the RG equation $\mathrm{d}\Delta_Z/\mathrm{d} l = (2-g_B)\Delta_Z(l)$, where the interaction dependent scaling dimension $g_B$ follows from the diagonalized Hamiltonian. Due to the presence of Coulomb repulsion, we find that $g_B$ is always smaller than its non-interacting value $g_0=1$, such that the gap is enhanced by hole-hole interactions. We can thus conclude that hole-hole interactions would stabilize a MBS in the presence of proximity-induced superconductivity, similar to proximitized Rashba NWs for electrons.\cite{Gangadharaiah2011,Stoudenmire2011} 
The RG flow is integrated until the running $\Delta_Z(l)$ grows to the value $\Delta_Z(l)/(\hbar v_{F_i}/\alpha)\sim1$,\cite{Giamarchi2003} signaling the opening of the helical gap. In physical units, this gap has the size $\Delta_Z^{*} = \Delta_Z^0\,(\hbar v_{F_i}/(2 \Delta_Z^0 \alpha))^{(1-g_B)/(2-g_B)}$.\cite{Braunecker2010}
In Fig.~\ref{fig:rescaledZeemangap} a), we plot both $\Delta_Z$ and $\Delta_Z^{*}$ as 
functions of $E_{\perp}$ for $B_{\perp} = 0.1\mbox{ T}$ and $B_z = 0 \mbox{ T}$, with $\alpha = 5.65 \mbox{ \AA}$ being the lattice constant of Ge.\cite{Winkler2003} 
We find that $\Delta_Z^{*}$ depends much stronger on $E_{\perp}$ than $\Delta_Z$. This can be attributed to the large changes in $v_{F_i}$ for decreasing $E_{\perp}$.  
In Fig.~\ref{fig:rescaledZeemangap} b), we finally display $\Delta_Z$ and $\Delta_Z^{*}$ for fixed $B$ and $E_{\perp}$ as functions of $\vartheta_B$, i.e.\ the direction of $\bm{B}$ with respect to the NW, and find that both $\Delta_Z$ and $\Delta_Z^{*}$ are strongly anisotropic. 
\begin{figure}[t]
\includegraphics[width = \columnwidth]{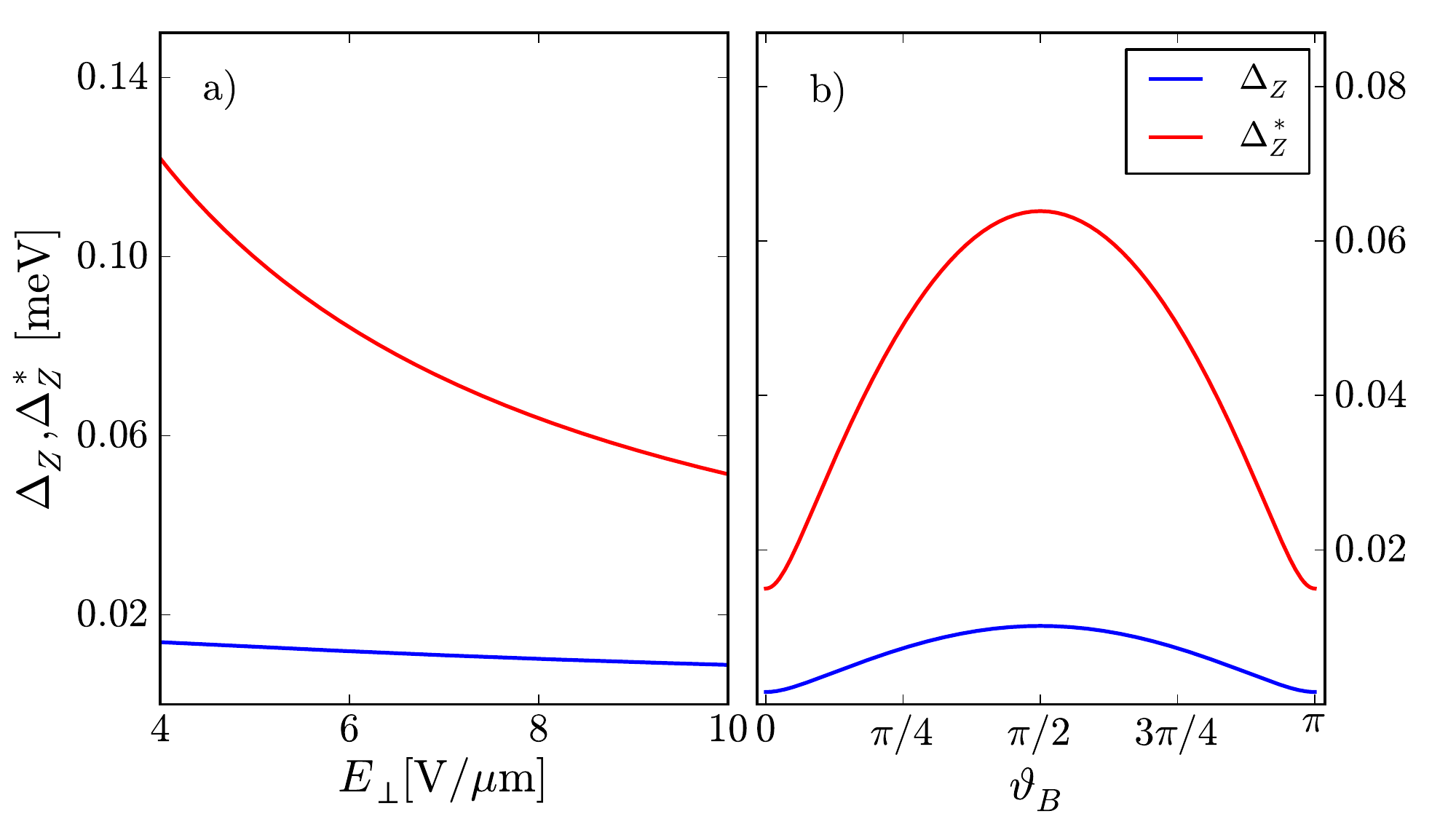}
\caption{The bare and rescaled helical gaps $\Delta_Z$ and $\Delta_Z^{*}$ for $B = 0.1\mbox{ T}$ (a) as functions of $E_{\perp}$ for $\vartheta_B = \pi/2$ and (b) as functions of $\vartheta_B$ for $E_{\perp} = 8 \mbox{ V/$\mu$m}$.  
We use the NW parameters $R = 7.5 \mbox{ nm}$ and $\Delta = 13\mbox{ meV}$ and fix $\varphi_E = 3\pi/2$ and $\bm{r}_{\text{mc}} = (0,|\bm{r}_{\text{mc}}|,0)$ with $|\bm{r}_{\text{mc}}|=100\mbox{ nm}$.}
\label{fig:rescaledZeemangap}
\end{figure}

\section{Conclusions}
In this work, we have addressed and quantified the effects of hole-hole interactions and their Luttinger liquid description in Ge/Si core/shell NWs, where we focused on the single subband regime most relevant for the emergence of MBS. We derived the Luttinger liquid description of the NW, and calculated the interaction dependent scaling exponents of various correlation functions. We showed a weak dependence of the scaling exponents on the magnitude of an applied electric field and a strong dependence on the NW parameters. 
Furthermore, the exponents revealed strong interaction effects since they differ substantially from their non-interacting value with thin NWs showing the strongest deviations.  To show the experimental relevance of our results, we analyzed the renormalization of the partial gap around zero momentum resulting from an applied magnetic field which is considerably enhanced by the hole-hole interactions. Additionally, we found that the gap is strongly anisotropic. Regarding the emergence of MBS in the helical regime of a superconducting hybrid device, the enhancement of the gap implies more stable MBS in an interacting system. 

In conclusion, hole-hole interactions show sizable effects in Ge/Si core/shell NWs and may lead to the stabilization of emerging MBS by enhancing the partial gap.

\begin{acknowledgments}
We thank Kevin van Hoogdalem and Christoph Kloeffel for useful discussions, and acknowledge funding from the Swiss NF, NCCR QSIT, and through the ECFP7-ICT initiative under
project SiSPIN No. 323841.
\end{acknowledgments}
%
%

%
\FloatBarrier
%
%
%
%
%
%
%
%
%
%
%

\appendix
\section{Calculation of the screened Coulomb matrix elements \label{app:calc_CoulMatel}}
To calculate the effective 1D, momentum dependent interaction matrix elements $V^{ijkl}_{1D}(q)$ introduced in Eq.~(\ref{eq:Hinteract}) of the main text, we use the transverse part of the real space wavefunctions of holes confined to the Ge core of the NW,\cite{Kloeffel2011} $\phi_{g_{\pm}}(\bm{r}_{\perp}) = g_{\pm}(r,\varphi)$ and $\phi_{e_{\pm}}(\bm{r}_{\perp}) = e_{\pm}(r,\varphi)$, where $\bm{r}_{\perp}$ denotes the transverse part of $\bm{r}$.
Due to the hard wall confinement assumed for the derivation of $\phi_{g_{\pm}}(\bm{r}_{\perp})$ and $\phi_{e_{\pm}}(\bm{r}_{\perp})$, the wavefunctions are proportional to functions of the type $J_n(r)$, where $J_n$ denotes the $n$th Bessel function of the first kind. 
The interaction matrix elements are given by integrals of form
\begin{align}
V^{ijkl}_{1D}(q) &= \int\!\!\!\!\int_{\text{NW}_{\perp}}\mathrm{d}\bm{r}_{\perp}\mathrm{d}\bm{r}'_{\perp}\phi_{i}^{\dagger}(\bm{r}_{\perp})\phi_{j}^{\dagger}(\bm{r}'_{\perp}) V(\bm{r},\bm{r}',\bm{r}_{\mathrm{mc}})\nonumber\\
&\phantom{=}\times\phi_{k}(\bm{r}'_{\perp})\phi_{l}(\bm{r}_{\perp}), \label{eq:interactionmatrixelement}
\end{align}
with $i,j,k,l = g_{\pm}, e_{\pm}$ and $\mathrm{d}\bm{r}_{\perp} = r\mathrm{d}r\mathrm{d}\varphi$ ($\mathrm{d}\bm{r}'_{\perp} = r'\mathrm{d}r'\mathrm{d}\varphi'$). 
Here, $\text{NW}_{\perp}$ indicates integration over the NW cross section. 
To perform the integration, we follow the procedure outlined in Ref.~[\onlinecite{Gold1990}].
We rewrite the summands of $V(\bm{r},\bm{r}',\bm{r}_{\mathrm{mc}})$ in terms of discrete Fourier transformations along the NW with wavevector $\bm{q} = (0,0,q)$, which brings the position dependent part of the Coulomb potential to the form
\begin{equation}
\frac{1}{|\bm{r}-\bm{r}'|} = \frac{1}{\sqrt{(z-z')^2+a^2}} =\sum_q 2 K_0(a q)e^{i q (z-z')},\label{eq:FTCoulomb}
\end{equation}
with $a^2 = r^2+r'^2-2 r r' \cos{(\varphi-\varphi')}$, and where $K_m(x)$ denotes the modified Bessel function of second kind. 
A similar term is obtained for position dependent part of the screened potential, $1/|\bm{r}-\bm{r}_{\text{mc}}|$.
Eq.~(\ref{eq:FTCoulomb}) can be simplified further by applying Graf's addition theorem for Bessel functions,\cite{Abramowitz1964} 
\begin{eqnarray}
K_0(a q)&=& \sum_{m=-\infty}^{\infty} e^{im (\varphi-\varphi')}K_m(q r_{>})I_m(q r_<), \label{eq:ResultGraf}
\end{eqnarray}
where $r,r' = r_{>},r_{<}$ with $r_{>}\geq r_<$, while $I_m(x)$ denotes the modified Bessel function of the first kind.
We insert the results of Eqs.~(\ref{eq:FTCoulomb}) and (\ref{eq:ResultGraf}) into Eq.~(\ref{eq:interactionmatrixelement}), and integrate out the angular part, $\int\!\!\!\int \mathrm{d}\varphi\mathrm{d}\varphi'$. 
The remaining non-zero contributions of the sum in Eq.~(\ref{eq:ResultGraf}) are the terms corresponding to $m=0,\pm1$.  
However, this result depends strongly on the exact form of the angular dependence of $\phi_{g_{\pm}}(\bm{r}_{\perp})$ and $\phi_{e_{\pm}}(\bm{r}_{\perp})$. 
The last step, the radial integration $\int\!\!\!\int r r' \mathrm{d}r \mathrm{d} r'$ cannot be performed directly in an analytical manner. 
To circumvent this, we replace the Bessel functions in Eq.~(\ref{eq:ResultGraf}), and in the wavefunctions by Taylor expansions around $r,r'=0$ up to appropriate order. 
This allows us to evaluate the radial integrals analytically. We have checked numerically that our analytical expressions reproduce the exact result very well.
%
%
%
%
%
%
%
\section{Operators and correlation functions in the $i,o$ basis, and transformation to the diagonal basis}\label{sec:app_op}

For the evaluation of correlation functions, it is helpful to change to a basis in which the matrices $\mathscr{H}_{\phi}$ and $\mathscr{H}_{\theta}$, given in the main text in Eqs.~(\ref{eq:mat1}) and (\ref{eq:mat2}), are diagonal.
This can be achieved by the basis change $(\phi_i,\phi_o)^T =W_{\phi} (\phi_p, \phi_m)^T$ and $(\theta_i,\theta_o)^T =W_{\theta} (\theta_p, \theta_m)^T$,
where $W_{\phi}$ and $W_{\theta}$ are matrices with (so far unspecified) real entries $w_{\phi,kl}$ and $w_{\theta,kl}$, and where we have introduced the new fields $\phi_{r}$ and $\theta_{r}$, with $r = p,m$.
For this transformation to be canonical, we demand that the new fields obey the commutation relations $\left[\phi_{r}(z),\nabla \theta_{r'}(z')\right] = i \,\pi \delta_{z,z'}\delta_{r,r'}$ and $\left[\phi_{r}(z), \theta_{r'}(z')\right] = i \,\pi/2\, \text{sgn}(z-z') \delta_{r,r'}$, which fixes $4$ of the $8$ parameters in $W_{\phi}$ and $W_{\theta}$.

The Hamiltonian densities are  transformed as $\tilde{\mathscr{H}}_{\phi} =(W_{\phi})^T \mathscr{H}_{\phi} W_{\phi}$ and $\tilde{\mathscr{H}}_{\theta} =(W_{\theta})^T \mathscr{H}_{\theta} W_{\theta}$. The requirement that the transformation diagonalizes the Hamiltonian densities fixes two more parameters in $W_{\phi}$ and $W_{\theta}$. The remaining two parameters are finally chosen such that

\begin{equation}
\tilde{\mathscr{H}}_{\phi} = \tilde{\mathscr{H}}_{\theta} = \frac{1}{2\pi}\left(\begin{array}{cc}u_p&0\\0&u_m\end{array}\right), \label{eq:Hamdens_diag}
\end{equation}
with velocities $u_p$ and $u_m$. 

This diagonal basis is particularly convenient if one is interested in evaluating correlation functions of the charge and spin degrees of freedom, where charge and spin are defined in analogy to a Rashba NW (the band $E_{g_+}'$, i.e.~the modes $L_o$ and $R_i$, are thus interpreted as left and right moving modes with spin up, while the band $E_{g_-}'$, i.e.~$L_i$ and $R_o$, are identified with left and right moving spin down modes). The operators describing the integrated charge ($\rho$) and spin ($\sigma$) densities ($\phi_i$) and currents ($\theta_i$) are given by
\begin{align}
&\phi_{\sigma} = -\frac{1}{\sqrt{2}}(\theta_i-\theta_o), &\theta_{\sigma} = -\frac{1}{\sqrt{2}}(\phi_i-\phi_o),\\  
&\phi_{\rho} = \frac{1}{\sqrt{2}}(\phi_i+\phi_o), &\theta_{\rho}= \frac{1}{\sqrt{2}}(\theta_i+\theta_o).
\end{align}
Using these relations, the operators for the charge density wave ($\text{CDW}$), spin density wave ($\text{SDW}$), singlet superconductivity ($\text{SS}$), and triplet superconductivity $(\text{TS}$) read 
\begin{align}
 O_{\text{CDW}}(r) &= \frac{1}{\pi \alpha} e^{- i k_{F_o}z }e^{i(\phi_{i}(r) + \phi_{o}(r))}\cos[\theta_{i}(r)-\theta_{o}(r)],\\
O_{\text{SDW}}^{x}(r) &= \frac{1}{\pi \alpha} e^{- i k_{F_o}z }e^{i(\phi_{i}(r) + \phi_{o}(r))}\nonumber\\
&\phantom{=}\times\cos[\phi_{i}(r)-\phi_{o}(r) + k_{F_o} z],\\
O_{\text{SDW}}^{y}(r) &= \frac{1}{\pi \alpha} e^{- i k_{F_o}z }e^{i(\phi_{i}(r) + \phi_{o}(r))}\nonumber\\
&\phantom{=}\times\sin[\phi_{i}(r)-\phi_{o}(r) + k_{F_o} z],\\
O_{\text{SDW}}^{z}(r) &= \frac{i}{\pi \alpha} e^{- i k_{F_o}z }e^{i(\phi_{i}(r) + \phi_{o}(r))}\sin[\theta_{o}(r)-\theta_{i}(r)],\\
O_{\text{SS}}(r)& = \frac{1}{\pi \alpha} e^{-i(\theta_{i}(r) + \theta_{o}(r))}\cos[\theta_{i}(r)-\theta_{o}(r)],\\
O_{\text{TS}}^{x}(r) &= \frac{1}{\pi \alpha} e^{-i(\theta_{i}(r) + \theta_{o}(r))}\cos[\phi_{i}(r)-\phi_{o}(r)\!+\! k_{F_o}z],\\
O_{\text{TS}}^{y}(r) &= \frac{1}{\pi \alpha} e^{-i(\theta_{i}(r) + \theta_{o}(r))}\sin[\phi_{i}(r)-\phi_{o}(r)\!+\! k_{F_o}z],\\
O_{\text{TS}}^{z}(r) &= \frac{i}{\pi \alpha} e^{-i(\theta_{i}(r) + \theta_{o}(r))}\sin[\theta_{o}(r)-\theta_{i}(r)]~.
\end{align}
A detailed discussion of these operators can be found in Ref.~[\onlinecite{Giamarchi2003}]. In the diagonal basis, the associated correlation functions are given by
\begin{align}
&\langle O_{\text{CDW}}^{\dagger}(r)O_{\text{CDW}}(0)\rangle = \langle O_{\text{SDW}}^{z,\dagger}(r)O_{\text{SDW}}^{z}(0)\rangle\nonumber\\
&\phantom{=}= \frac{1}{(\pi \alpha)^2} e^{i k_F z}(\alpha/r)^{1/2(u_{a+}^2+ u_{b+}^2+ w_{a-}^2+ w_{b-}^2)}, \\
\nonumber\\
&\langle O_{\text{SDW}}^{x,\dagger}(r)O_{\text{SDW}}^{x}(0)\rangle = \langle O_{\text{SDW}}^{y,\dagger}(r)O_{\text{SDW}}^{y}(0)\rangle\nonumber\\
&\phantom{=}= \frac{1}{4(\pi \alpha)^2} \left[e^{2 i k_F z}
(\alpha/r)^{1/2[(u_{a+}- u_{a-})^2+ (u_{b+}- u_{b-})^2]}\right.\nonumber\\
&\phantom{===}\left.+ (\alpha/r)^{1/2[(u_{a+}+ u_{a-})^2+ (u_{b+}+ u_{b-})^2]} \right],\nonumber\\
&\phantom{=}= \langle O_{\text{SDW}}^{y,\dagger}(r)O_{\text{SDW}}^{y}(0)\rangle_{(2 k_F)} + \langle O_{\text{SDW}}^{y,\dagger}(r)O_{\text{SDW}}^{y}(0)\rangle_{(0)} \\
\nonumber\\
&\langle O_{\text{SS}}^{\dagger}(r)O_{\text{SS}}(0)\rangle = \langle O_{\text{TS}}^{z,\dagger}(r)O_{\text{TS}}^{z}(0)\rangle\nonumber\\
&\phantom{=}= \frac{1}{4(\pi \alpha)^2} \left[
(\alpha/r)^{1/2[(w_{a+}+ w_{a-})^2+ (w_{b+}+ w_{b-})^2]} \right.\nonumber\\
&\phantom{===}\left.+(\alpha/r)^{1/2[(w_{a+}- w_{a-})^2+ (w_{b+}- w_{b-})^2]} \right],\nonumber\\
&\phantom{=}= \langle O_{\text{TS}}^{z,\dagger}(r)O_{\text{TS}}^{z}(0)\rangle_{(+)}+ \langle O_{\text{TS}}^{z,\dagger}(r)O_{\text{TS}}^{z}(0)\rangle_{(-)}\\
\nonumber\\
&\langle O_{\text{TS}}^{x,\dagger}(r)O_{\text{TS}}^{x}(0) = \langle O_{\text{TS}}^{y,\dagger}(r)O_{\text{TS}}^{y}(0)\rangle\nonumber\\
&\phantom{=}= \frac{1}{(\pi \alpha)^2} (\alpha/r)^{1/2(u_{a-}^2+ u_{b-}^2+ w_{a+}^2+ w_{b+}^2)}, 
\end{align}
with $u_{a,\pm} = w_{\phi,21}\pm w_{\phi,11}$, $u_{b,\pm} = w_{\phi,22}\pm w_{\phi,21}$, $w_{a,\pm} = w_{\theta,21}\pm w_{\theta,{11}}$, $w_{b,\pm} = w_{\theta,22}\pm w_{\theta,{12}}$. 
Note that the correlation functions for $\text{SDW}_{x,y}$, $\text{SS}$, and $\text{TS}_z$ each contain two terms with (slightly) different exponents.
%
%
%
%
%
%
%
%
%
%

\section{Divergences outside the perturbative regime and comparison to a fermionic RG approach\label{app:divergence_exps}}
As discussed in the main text, the use of a perturbative RG approach for the coupling $g_I$ restricts our analysis to a regime in which the associated dimensionless coupling satisfies $2 g_I/(u_p+u_m)\ll 1$. In this regime, $g_I$ is found to be RG irrelevant. When leaving the perturbative regime, we find that the scaling exponents diverge at a finite field $E_{\perp}$ (solid lines in Fig.~\ref{fig:exponents_SG_app}). These divergences can be traced back to the fact that the off-diagonal matrix elements in the Hamiltonian given in Eq.~\eqref{eq:bos_mat} are so large that the velocity $u_m$ of the diagonalized Hamiltonian vanishes. The system thus seems to exhibit a Wentzel-Bardeen singularity.\cite{Wentzel1951, Bardeen1951, Loss1994,Martin1995} This apparent divergence, however, occurs outside the perturbative regime, and is thus beyond the range of validity of our calculation (we note that during the RG flow, the quadratic sector of the theory is renormalized by corrections of the order $\mathcal{O}(g_I^2)$, 
see Ref.~[\onlinecite{Giamarchi2003}]).

As an independent cross-check for the absence of singularities in the regime described by our calculation, we compare the scaling exponents of the correlation functions discussed in the main text to the analogous exponents obtained when the system is bosonized only after an initial fermionic one-loop RG treatment, which in particular  already describes the flow of $g_I$ to weak coupling. To this end, we start from the fermionic Hamiltonian with the interactions given in Eqs.~\eqref{eq:H2} - \eqref{eq:H1}, and perform a fermionic one-loop RG analysis following Refs.~[\onlinecite{Muttalib1986}] and [\onlinecite{Schulz2009}]. This yields the one-loop RG equations
\begin{equation}
\frac{\mathrm{d} \bar{g}_2}{\mathrm{d} l} = -\bar{g}_I(l)^2, \qquad
\frac{\mathrm{d} \bar{g}_I}{\mathrm{d} l} = -\bar{g}_I(l) \bar{g}_2(l),
\end{equation}
where we have introduced the definitions $ \pi \bar{g}_2 =  g_{2_o}/(2  v_{F_o})+ g_{2_i}/(2 v_{F_i})- 2(g_{2_{io}}-g_{1_{io1}})/(v_{F_o}+v_{F_i})$ and $\pi \bar{g}_I =  g_I\sqrt{2(1+\gamma)} / (v_{F_o}+v_{F_i})$, with $\gamma = (v_{F_o}+v_{F_i})^2/(4 v_{F_o} v_{F_i})$. Integrating these RG equations yields the fixed point values $\bar{g}_I^{*} =  0$ and $\bar{g}_2^{*} = (\bar{g}_2^2(0)-\bar{g}_I^2(0))^{1/2}$,\cite{Giamarchi2003} from which we find $g_{2_o}^{*}$, $g_{2_i}^{*}$ and $g_{2_{io}}^{*}-g_{1_{io}}^{*}$. We plug these renormalized interactions back into the fermionic Hamiltonian. The bosonization of this renormalized Hamiltonian in turn yields a purely quadratic bosonic theory, which finally allows to calculate the exponents of the various correlation functions just as discussed in the main text.

In the perturbative regime, we find that the qualitative behavior of the scaling exponents is identical for both approaches. Outside the perturbative regime, this is not true. There, the presence or absence of a divergence of the scaling exponents calculated using the fermionic one-loop approach depends on the NW parameters, while the bosonic approach seems to generically exhibit a singularity.

Most importantly, however, we find that - if present - the divergences always appear outside the range of validity of our calculation. This finding is illustrated in Fig.~\ref{fig:exponents_SG_app}, where we plot the exponents obtained in the bosonic sine-Gordon approach (solid lines), and the exponents obtained when bosonizing after the fermionic one-loop RG treatment (dashed lines). 
The black vertical lines denote the limits of the perturbative regime, \textit{i.e.}~the black solid line indicates where 
\begin{equation}
\frac{g_I}{\frac{1}{2}(u_p + u_m)} \approx 1~,
\end{equation}
while the black dashed line depicts where 
\begin{equation}
\bar{g}_2=\frac{g_{2_o}}{2 \pi  v_{F_o}}+\frac{ g_{2_i}}{2\pi v_{F_i}}- \frac{2(g_{2_{io}}-g_{1_{io1}})}{\pi(v_{F_o}+v_{F_i})}\approx 1~.
\end{equation}
\begin{figure}[t]
\includegraphics[width =\columnwidth]{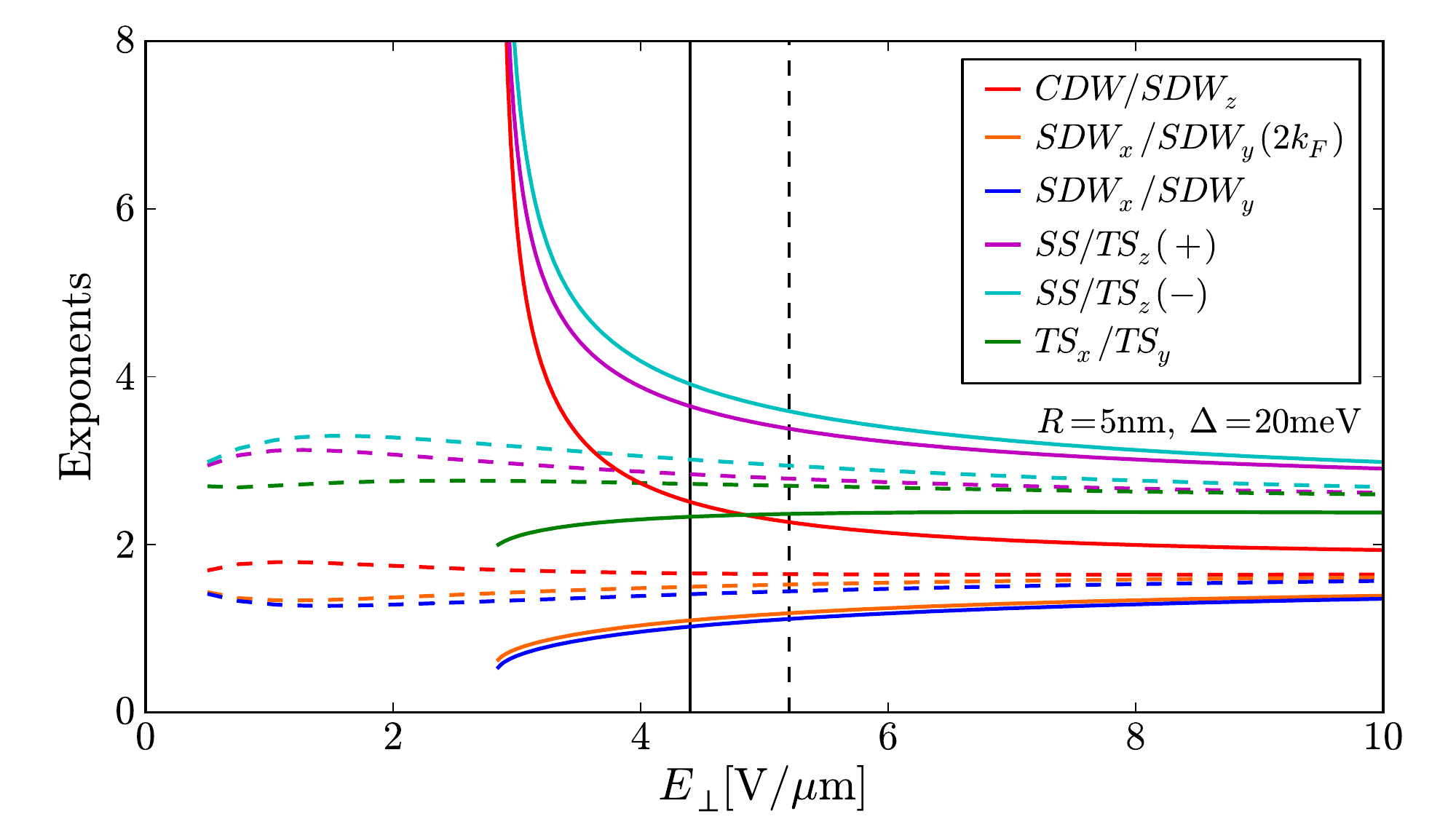}
\includegraphics[width =\columnwidth]{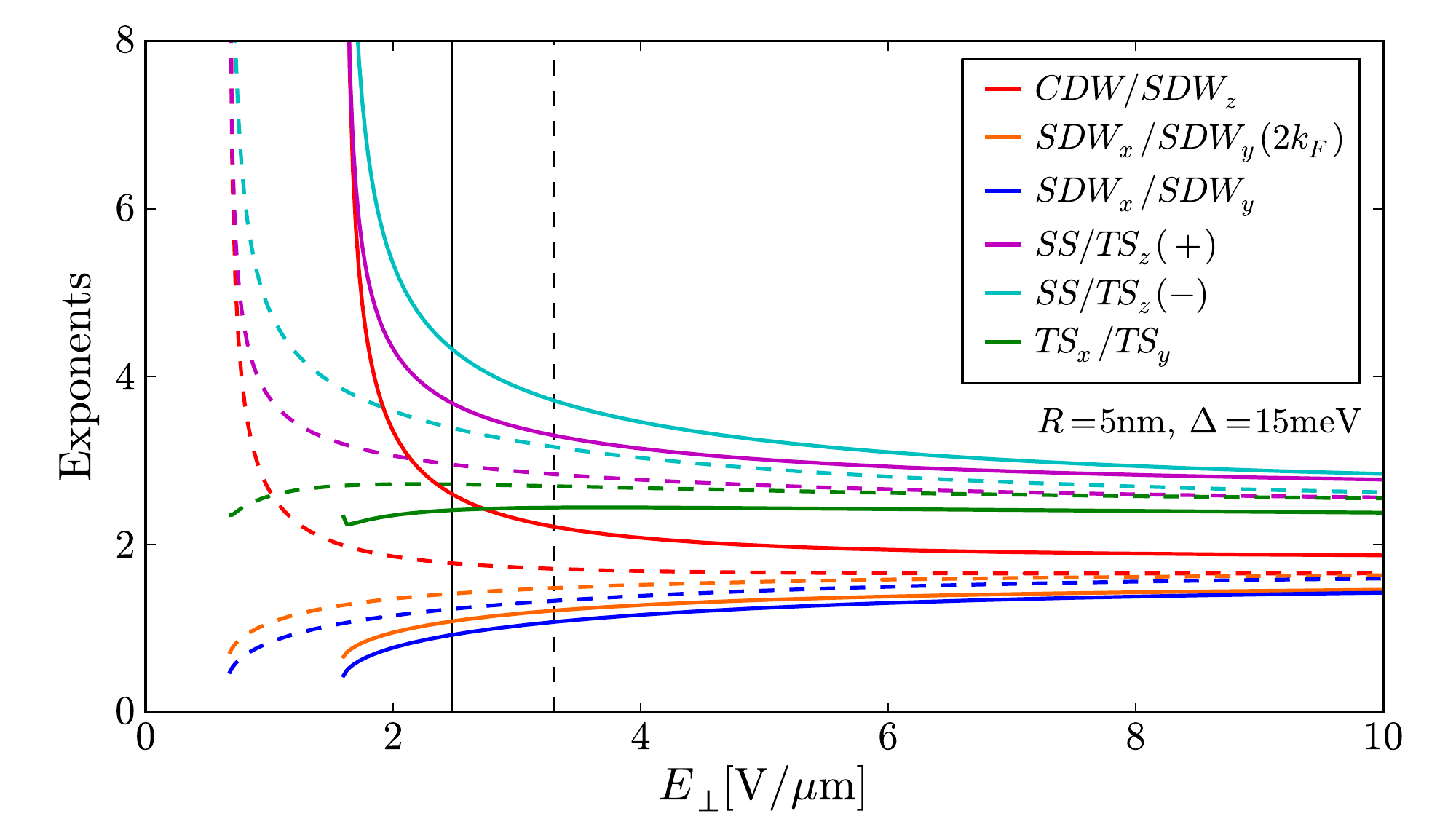}
\caption{
The exponents of the correlation functions for two different sets of NW parameters $R$ and $\Delta$ as functions of $E_{\perp}$, where $\varphi_E = 3\pi/2$ and $\bm{r}_{\text{mc}} = (0,|\bm{r}_{\text{mc}}|,0)$ with $|\bm{r}_{\text{mc}}|=100\mbox{ nm}$. 
We show the exponents obtained by two different RG approaches,  a bosonic sine-Gordon RG approach (solid lines), and a fermionic one-Loop RG calculation (dashed lines). 
The black vertical lines denote where the perturbative regime ends for either of the two RG approaches. 
In contrast to the bosonic approach, where the scaling exponents always diverge, the presence of a divergence of the scaling exponents for the fermionic approach depends on the NW parameters. 
\label{fig:exponents_SG_app}}
\end{figure}

\FloatBarrier


%

\end{document}